%% file: main.tex
\documentclass{ifacconf}

\usepackage{graphicx}      
\usepackage{natbib}        

\usepackage{amsmath} 
\usepackage{amssymb}  
\usepackage{bm}
\usepackage{mathtools}
\usepackage{tabularx}
\usepackage{booktabs}
\usepackage{multirow}
\usepackage{units}
\usepackage[super]{nth}
\allowdisplaybreaks
\usepackage{tikz}
\usepackage{pgfplots}
\usepackage{tikzscale}
\pgfplotsset{compat = 1.3}

\usepackage{algorithm}
\usepackage[noend]{algpseudocode}

\usepackage{balance}
\usepackage{url}

\newcommand{\norm}[1]{\left\lVert#1\right\rVert}
\newcommand{\diag}{\mathrm{diag}}

\newtheorem{assumption}{Assumption}
\newtheorem{remark}{Remark}
\newtheorem{definition}{Definition}

\def\ct-#1{ct\nobreakdash-#1}
\def\RPI-#1-#2{RPI\nobreakdash-#1\nobreakdash-#2}
\def\SL-#1{SL\nobreakdash-#1}
\def\tube-#1{tube\nobreakdash-#1}

\newcolumntype{C}{>{\centering\arraybackslash}X}

\makeatletter
\newcommand{\eqnum}{\refstepcounter{equation}\textup{\tagform@{\theequation}}}
\makeatother

\usepackage{color}

\newif\ifcommentandrea
\commentandreafalse

\begin{document}
\begin{frontmatter}

\title{Asynchronous Computation of Tube-based Model Predictive Control\thanksref{footnoteinfo}} 

\thanks[footnoteinfo]{This work was supported by the European Space Agency (ESA) under NPI 621-2018 and the Swiss Space Center (SSC).}

\author[First]{Jerome Sieber} 
\author[First]{Andrea Zanelli} 
\author[First]{Antoine P. Leeman} 
\author[Second]{Samir Bennani}
\author[First]{Melanie N. Zeilinger} 

\address[First]{Institute for Dynamic Systems and Control, ETH Zurich, 8092 Zurich, Switzerland (e-mail: \{jsieber,zanellia,aleeman,mzeilinger\}@ethz.ch)}
\address[Second]{ESA-ESTEC, Noordwijk 2201 AZ, The Netherlands\\(e-mail: samir.bennani@esa.int).}

\begin{abstract}                
Tube-based model predictive control (MPC) methods bound deviations from a nominal trajectory due to uncertainties in order to ensure constraint satisfaction. While techniques that compute the tubes online reduce conservativeness and increase performance, they suffer from high and potentially prohibitive computational complexity. This paper presents an asynchronous computation mechanism for system level tube-MPC (SLTMPC), a recently proposed tube-based MPC method which optimizes over both the nominal trajectory and the tubes. Computations are split into a primary and a secondary process, computing the nominal trajectory and the tubes, respectively. This enables running the primary process at a high frequency and moving the computationally complex tube computations to the secondary process. We show that the secondary process can continuously update the tubes, while retaining recursive feasibility of the primary process.
\end{abstract}

\begin{keyword}
Predictive control,
Uncertain systems,
Robust control (linear case)
\end{keyword}

\end{frontmatter}

\section{Introduction}
Tube-based model predictive control~(MPC) methods are the principal robust control techniques for constrained linear systems affected by bounded additive disturbances and bounded model uncertainties. These methods are based on separating the system behavior into nominal and error dynamics, which describe the system dynamics neglecting uncertainties and the deviations from the nominal dynamics due to uncertainties, respectively. Since the uncertainties are assumed to be bounded, the error trajectories can be bounded in so-called tubes, which are computed based on the uncertainty bound and a tube controller. Tube-based MPC methods compute the tubes offline and optimize the nominal trajectory subject to tightened constraints online, see e.g.,~\citep{Chisci2001,Mayne2005,Zanon2021}. In order to reduce conservativeness, there has been increased interest in more flexible tube-based formulations, which also allow online optimization of the tubes, e.g.,~\citep{Rakovic2012b,Rakovic2016} present homothetic and elastic tubes, which are fixed in shape, but can be dilated and translated online. The system level tube-MPC (SLTMPC) approach introduced in~\citep{Sieber2021}, computes the tubes fully online by optimizing the tube controller. However, if both the nominal trajectory and the tubes are optimized online, the optimization problem becomes computationally expensive. To address this issue, we propose an asynchronous computation mechanism for tube-based MPC methods.
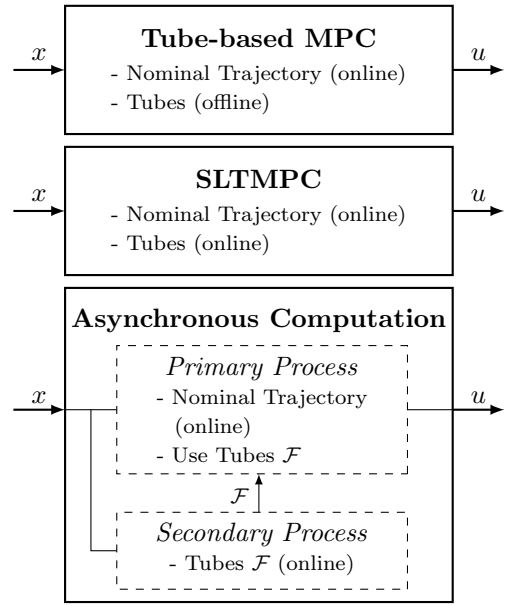
\begin{figure}[t]
\centering
\input{tikz/comparison}
\caption{Comparison between tube-based MPC, SLTMPC, and tube-based MPC with asynchronous computation, separating the nominal trajectory and tube computations into different processes.}
\label{fig:architecture} 
\end{figure}
Specifically, we introduce the mechanism for SLTMPC, however, the same principle can also be applied to other tube-based MPC methods that compute the tubes online, e.g., configuration-constrained tubes proposed in~\citep{Villanueva2022}. Figure~\ref{fig:architecture} compares tube-based MPC, SLTMPC, and the proposed asynchronous computation mechanism in terms of their offline and online computations. The main idea of the asynchronous computation mechanism is to split nominal trajectory and tube computations into separate processes instead of computing them simultaneously. The \emph{primary process} computes the nominal trajectory, while the \emph{secondary process} computes the tubes, which are then passed to the primary process. This brings the computational complexity of the primary process closer to that of tube-based MPC, which allows running the SLTMPC controller at higher frequencies, while retaining the benefits of online optimized tubes.

\textit{Contributions:}
We consider SLTMPC applied to linear time-invariant dynamical systems with additive uncertainties. For this class of controllers, we introduce the asynchronous computation mechanism and explain how nominal trajectory and tube computations are split into separate processes. We propose a primary process that retains recursive feasibility when asynchronously updating the tubes by \emph{fusing} old and new tubes based on ideas from~\citep{Kogel2020b}. Additionally, we present a new type of scaled terminal set for SLTMPC methods.

\textit{Related Work:}
The system level parameterization~(SLP), which forms the basis of SLTMPC, was introduced in~\citep{Anderson2019}. The SLP was primarily used in the context of distributed systems, see e.g.,~\citep{Alonso2019}, before it was applied to tube-based MPC in~\cite{Sieber2021}.
A similar concept to asynchronous computation is used in multi-level real-time iteration algorithms, where computations are performed in parallel using a hierarchical structure, see e.g.,~\citep{Bock2007}. This paper is closest related to~\citep{Kogel2020b,Kogel2020a}, which introduce the idea of fusing and blending tubes online. We extend this idea to system level tubes to formulate the primary process.

The remainder of the paper is organized as follows: Section~\ref{sec:preliminaries} introduces the notation, basic definitions, and tube-based MPC formulations. In Section~\ref{sec:asynch-up} we discuss the general system level tube-MPC (SLTMPC) problem, before introducing SLTMPC with asynchronous computation. Finally, Section~\ref{sec:numerical_section} presents a numerical example and Section~\ref{sec:conclusions} concludes the paper.

\section{Preliminaries}\label{sec:preliminaries}
\subsection{Notation and Basic Definitions}
In this paper, we denote vectors with lower case letters, e.g.,~$a$, and matrices with upper case letters, e.g.,~$A$. With bold letters we indicate stacked column vectors, e.g.,~$\mathbf{a}$, and block matrices, e.g.,~$\mathbf{A}$, whose elements we index with superscripts, e.g.,~$a^i, \ A^{i,j}$. We denote the space of symmetric positive definite matrices as $\mathbb{S}^n_{++} = \{ A \in \mathbb{R}^{n \times n} \!\mid\! A=A^\top\!, A \succ 0\}$. For two vectors~$x, \, y$, we use the shorthand~$(x, y)$ to denote $[x^\top, y^\top]^\top$ and for vector~$x$ and $P \in \mathbb{S}_{++}$ we use~$\norm{x}^2_P$ to denote the quadratic form $x^\top P x$. Additionally, we use~$\diag(A)^i$ to denote the block-diagonal matrix, whose diagonal consists of $i$-times the $A$ matrix. We distinguish between the states of a dynamical system~$x(k)$ and the states predicted by an MPC algorithm~$x_i$. We use~$\oplus$ and~$\ominus$ to denote Minkowski set addition and Pontryagin set subtraction, respectively, which are both formally defined in~\cite[Definition~3.10]{Rawlings2009}. For a sequence of sets~$\mathcal{A}_0, \dots, \mathcal{A}_j$, we use the convention that~$\bigoplus_{i=0}^{-1} \mathcal{A}_i = \{0\}$, which is the set containing only the zero vector.
Below, we state the definitions of robust set invariance~\cite[]{Blanchini99}, disturbance reachable sets~\cite[]{Chisci2001}, and support functions~\cite[]{Kolmanovsky1998}, which we use throughout the paper.
\begin{definition}[Robust positively invariant (RPI) set]
A\\set $\mathcal{S} \subseteq \mathbb{R}^n$ is a RPI set for the system $x^+ = Ax + w, \, A \in \mathbb{R}^{n \times n}$, if $x^+ \in \mathcal{S}$ for all $x \in \mathcal{S}$, $w \in \mathcal{W} \subseteq \mathbb{R}^n$.
\end{definition}
\begin{definition}[Disturbance reachable sets (DRS)]
For the system $x^+ = Ax + w, \, A \in \mathbb{R}^{n \times n},\, w \in \mathcal{W} \subseteq \mathbb{R}^n$, where~$\mathcal{W}$ is a compact set containing the origin in its interior, the DRS are defined as~$\mathcal{F}_i = \bigoplus_{j=0}^{i-1}A^j\mathcal{W}, \, i \geq 0$.
\end{definition}
\begin{definition}[Support function]\label{def:sup_fn}
The support function of a non-empty compact convex set $\mathcal{A} \subseteq \mathbb{R}^n$, evaluated at $\eta \in \mathbb{R}^n$ is given by $h_\mathcal{A}(\eta) = \sup_{a \in \mathcal{A}} \eta^\top a$.
\end{definition}

\subsection{Tube-based Model Predictive Control}
We consider linear time-invariant dynamical systems with additive disturbances of the form
\begin{equation}\label{eq:dynamics}
x(k\!+\!1) = A x(k) + B u(k) + w(k),
\end{equation}
with $A \in \mathbb{R}^{n \times n}$, $B \in \mathbb{R}^{n \times m}$, and $w(k) \in \mathcal{W}$, where $\mathcal{W} \subseteq \mathbb{R}^n$ is a compact convex set containing the origin in its interior.
The system is subject to compact polytopic state and input constraints
\begin{subequations}\label{eq:constraints}
\begin{align}
\mathcal{X} &= \{ x \in \mathbb{R}^n \mid H_{x} x \leq h_{x}, \, H_{x} \!\in\! \mathbb{R}^{n_x \times n}, \, h_{x} \!\in\! \mathbb{R}^{n_x}\}, \\
\mathcal{U} &= \{ u \in \mathbb{R}^m \mid H_{u} u \leq h_{u}, \, H_{u} \!\in\! \mathbb{R}^{n_u \times m}, \, h_{u} \!\in\! \mathbb{R}^{n_u}\},
\end{align}
\end{subequations}
containing the origin in their interior. To control system~\eqref{eq:dynamics}, we consider MPC with horizon length~$N$. Therefore, we define $\mathbf{x} = ( x_0, \dots, x_N )$, $\bm{\delta} = ( x_0, \mathbf{w} ) = ( x_0,w_0, \dots, w_{N-1} )$, and $\mathbf{u} = ( u_0, \dots, u_N )$ as the state, disturbance, and input trajectories,\!\footnote{Note that we include the input~$u_N$ here for ease of notation. However, this input will be ignored in the MPC formulations.}respectively. Then, the dynamics are compactly written in terms of these trajectories as
\begin{equation}\label{eq:stacked_dynamics}
\mathbf{x} = \mathbf{ZA}_{N+1}\mathbf{x} + \mathbf{ZB}_{N+1}\mathbf{u} + \bm{\delta},
\end{equation}
where $\mathbf{Z}$ is a matrix with non-zero elements only on the $(n \!+\! 1)$-th sub-diagonal, and $\mathbf{A}_{N+1} = \diag(A)^{N+1}$, $\mathbf{B}_{N+1} = \diag(B)^{N+1}$. In this paper, we focus on tube-based MPC methods, which use control policies of the form $\mathbf{u} = \mathbf{v} + \mathbf{K}(\mathbf{x} - \mathbf{z})$, where $\mathbf{z} = ( z_0, \dots, z_N )$ and $\mathbf{v} = ( v_0, \dots, v_N )$ are the nominal state and input trajectories, respectively, and $\mathbf{K} \in \mathbb{R}^{(N+1)m\times (N+1)n}$ is a block-lower-triangular matrix and denotes the tube controller. Due to the linearity of the system, dynamics~\eqref{eq:stacked_dynamics} can be split into nominal and error dynamics, i.e.,
\begin{align}
\mathbf{z} &= \mathbf{ZA}_{N+1}\mathbf{z} + \mathbf{ZB}_{N+1}\mathbf{v} + \mathbf{z}_\mathbf{0}, \label{eq:nominal_dynamics} \\
\mathbf{x^e} &= \left(\mathbf{ZA}_{N} + \mathbf{ZB}_{N}\mathbf{K}\right)\mathbf{x^e} + \mathbf{w}, \label{eq:error_dynamics}
\end{align}
where $\mathbf{x^e} = ( x^e_1, \dots, x^e_N )$ is the error state, $\mathbf{z}_\mathbf{0} = (z_0,0,\dots,0)$, and $\mathbf{w} = (w_0, \dots, w_{N-1})$. This allows us to formulate tube-based MPC methods with respect to the nominal trajectories $\mathbf{z}, \mathbf{v}$ instead of the system trajectories $\mathbf{x}, \mathbf{u}$, while guaranteeing constraint satisfaction through tightened state and input constraints via bounds on the error dynamics~\eqref{eq:error_dynamics}, i.e., the tubes. For a more detailed treatment and a broader overview of tube-based MPC methods, see e.g.,~\citep{Rawlings2009}.

\section{System Level Tube-MPC with Asynchronous Computation}\label{sec:asynch-up}
In the following, we first introduce system level tube-MPC~(SLTMPC)~\citep{Sieber2021}, for which we then propose an asynchronous computation mechanism.

\subsection{System Level Tube-MPC (SLTMPC)}\label{sec:SLTMPC}
We consider SLTMPC, which uses the system level parameterization~(SLP)~\citep{Anderson2019} to parameterize error dynamics~\eqref{eq:error_dynamics} in terms of the \emph{error system responses} $\bm{\Phi}_\mathbf{x^e} \in \mathbb{R}^{Nn\times Nn}, \bm{\Phi}_\mathbf{u^e} \in \mathbb{R}^{Nm\times Nn}$~(compare also~\citep{Leeman2023}) as
\begin{equation*}
\mathbf{x^e} = \bm{\Phi}_\mathbf{x^e}\mathbf{w},\quad
\mathbf{u^e} = \mathbf{K}\mathbf{x^e} = \bm{\Phi}_\mathbf{u^e}\mathbf{w},
\end{equation*}
with $\mathbf{u^e} = ( u^e_1, \dots, u^e_N )$ and where $\bm{\Phi}_\mathbf{x^e}, \, \bm{\Phi}_\mathbf{u^e}$ are block-lower-triangular matrices and are defined as
\begin{align*}
\bm{\Phi}_\mathbf{x^e} \!=\! \left(I  \scalebox{0.8}[1.0]{\( - \)} \mathbf{ZA}_N  \scalebox{0.8}[1.0]{\( - \)} \mathbf{ZB}_N\mathbf{K}\right)^{ \scalebox{0.5}[1.0]{\( - \)}1}\!\!\!, \
\bm{\Phi}_\mathbf{u^e} \!=\! \mathbf{K}\!\left(I  \scalebox{0.8}[1.0]{\( - \)} \mathbf{ZA}_N  \scalebox{0.8}[1.0]{\( - \)} \mathbf{ZB}_N\mathbf{K}\right)^{ \scalebox{0.5}[1.0]{\( - \)}1}\!\!\!.
\end{align*}
Then, the state and input trajectories are defined as the sum of the nominal and error trajectories, i.e.,
\begin{subequations}
\begin{align*}
\mathbf{x} = \mathbf{z} \!+\! \begin{bmatrix} 0 \\ \mathbf{x^e}\end{bmatrix} = \mathbf{z} \!+\! \begin{bmatrix} 0 \\ \bm{\Phi}_\mathbf{x^e}\mathbf{w}\end{bmatrix}\!, \
\mathbf{u} = \mathbf{v} \!+\! \begin{bmatrix} 0 \\ \mathbf{u^e}\end{bmatrix} = \mathbf{v} \!+\! \begin{bmatrix} 0 \\ \bm{\Phi}_\mathbf{u^e}\mathbf{w}\end{bmatrix}\!.
\end{align*}
\end{subequations}
Using~\cite[Theorem~2.1]{Anderson2019} we can guarantee that the error system responses $\bm{\Phi}_\mathbf{x^e}, \bm{\Phi}_\mathbf{u^e}$ parameterize all error trajectories $\mathbf{x^e}, \mathbf{u^e}$ that are realized by tube controller $\mathbf{K}$ and we can equivalently write~\eqref{eq:error_dynamics} as
\begin{equation}\label{eq:SLP-affine-constraint}
\begin{bmatrix} I - \mathbf{ZA}_N & -\mathbf{ZB}_N \end{bmatrix} \!\begin{bmatrix} \bm{\Phi}_\mathbf{x^e} \\ \bm{\Phi}_\mathbf{u^e} \end{bmatrix} = I.
\end{equation}
Thereby, the error dynamics are completely defined by the closed-loop behavior, which allows optimization over system responses instead of the tube controller. Throughout this paper, we impose the following Toeplitz structure on the error system responses
\begin{equation}\label{eq:Toeplitz-structure}
\bm{\Phi}_\mathbf{x^e} \!=\!\! \begin{bmatrix} \Phi_{x^e}^{1} & & & \\ \Phi_{x^e}^{2} & \hspace*{-0.1cm}\Phi_{x^e}^{1} &  &  \\ \vdots & \hspace*{-0.1cm}\ddots & \hspace*{-0.1cm}\ddots & \\ \Phi_{x^e}^{N} & \hspace*{-0.15cm}\dots & \hspace*{-0.1cm}\Phi_{x^e}^{2} & \hspace*{-0.05cm}\Phi_{x^e}^{1} \end{bmatrix}\!\!, \,
\bm{\Phi}_\mathbf{u^e} \!=\!\! \begin{bmatrix} \Phi_{u^e}^{1} & & & \\ \Phi_{u^e}^{2} & \hspace*{-0.1cm}\Phi_{u^e}^{1} &  &  \\ \vdots & \hspace*{-0.1cm}\ddots & \hspace*{-0.1cm}\ddots & \\ \Phi_{u^e}^{N} & \hspace*{-0.1cm}\dots & \hspace*{-0.1cm}\Phi_{u^e}^{2} & \hspace*{-0.1cm}\Phi_{u^e}^{1} \end{bmatrix}\!\!.
\end{equation}
Using the error system responses $\bm{\Phi}_\mathbf{x^e}, \bm{\Phi}_\mathbf{u^e}$, the SLTMPC problem is defined as
\begin{subequations}\label{SLTMPC:generic}
        \begin{alignat}{2}
                \min_{\mathbf{z}, \mathbf{v}, \bm{\Phi}_\mathbf{x^e}, \bm{\Phi}_\mathbf{u^e}} \quad & \norm{z_N}^2_P + \sum_{i=0}^{N-1} \norm{z_i}^2_Q + \norm{v_i}^2_R \label{SLTMPC:cost}\\
                \textrm{s.t. } \qquad\, & \mathbf{z} = \mathbf{ZA}_{N+1}\mathbf{z} + \mathbf{ZB}_{N+1}\mathbf{v} + \mathbf{z}_\mathbf{0}, \label{SLTMPC:dynamics}\\
                & z_i \in \mathcal{X} \ominus \mathcal{F}_i^x, &&\hspace*{-1.4cm} i=0, \dots, N\!-\!1,\\
                & v_i \in \mathcal{U} \ominus \mathcal{F}_i^u, &&\hspace*{-1.4cm} i=0, \dots, N\!-\!1, \\
                & z_N \in \mathcal{Z}_f \subseteq \mathcal{X} \ominus \mathcal{F}_N^x, \\
                & \left( \bm{\Phi}_\mathbf{x^e}, \bm{\Phi}_\mathbf{u^e} \right) \in \mathcal{S}, \label{SLTMPC:structural-constraint}\\
                & z_0 = x(k), \label{SLTMPC:init}
        \end{alignat}
\end{subequations}
where $Q \in \mathbb{S}^n_{+}, \, R \in \mathbb{S}^m_{++}$ are the state and input cost parameters, $\mathcal{F}^x_i$, $\mathcal{F}^u_i$ are the state and input tubes, $\mathcal{S}$ is a set of structural constraints on the error system responses, and $\mathcal{Z}_f, \, P$ are suitable terminal ingredients, whose properties we discuss in the following.

The tubes $\mathcal{F}_i^x, \, \mathcal{F}_i^u$ are system level disturbance reachable sets (SL-DRS) and are defined in terms of the error system responses as
\begin{align*}
\mathcal{F}_i^x &= \bigoplus_{j=1}^{i} \Phi_{x^e}^{j} \mathcal{W}, \qquad \mathcal{F}_i^u = \bigoplus_{j=1}^{i} \Phi_{u^e}^{j} \mathcal{W},
\end{align*}
for which we use the shorthands $\mathcal{F}^x = \{ \mathcal{F}_0^x, \dots, \mathcal{F}_N^x\}$ and~$\mathcal{F}^u = \{ \mathcal{F}_0^u, \dots, \mathcal{F}_N^u\}$.
By definition, the SL-DRS admit the recursions
\begin{align} \label{SLDRS:inclusion}
\mathcal{F}_{i+1}^x &= \mathcal{F}_{i}^x \oplus \Phi_{x^e}^{i+1} \mathcal{W}, \qquad
\mathcal{F}_{i+1}^u = \mathcal{F}_{i}^u \oplus \Phi_{u^e}^{i+1} \mathcal{W}. 
\end{align}
The set of structural constraints $\mathcal{S}$ is defined as

\begin{figure}[t]
\centering
\input{tikz/asynchronous-updates}
\caption{Visualization of SLTMPC with asynchronous computation: the secondary process computes the tubes $\mathcal{F}^x, \, \mathcal{F}^u$ and terminal set $\alpha\mathcal{X}_f$, which are then passed to the primary process and stored in memory $\mathtt{M}$.}
\label{fig:asynch-up} 
\end{figure}
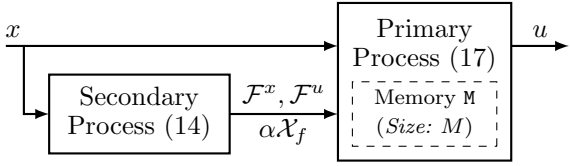

\vspace*{-0.2cm}
\begin{minipage}[]{0.28\columnwidth}
\begin{equation*}
\mathcal{S} \!=\! \left \lbrace \!(\bm{\Phi}_\mathbf{x^e}, \bm{\Phi}_\mathbf{u^e}) \!\;\middle|\; \begin{array}{l} \\[1cm] \end{array} \right.
\end{equation*}
\end{minipage}
\begin{minipage}[]{0.64\columnwidth}\vspace*{-0.1cm}
\begin{align}
        &\Phi_{x^e}^{i+1,j} \!= \Phi_{x^e}^{i,j}, \, \Phi_{u^e}^{i+1,j} \!= \Phi_{u^e}^{i,j}, \forall j, \label{eq:structure2} \\
        &\Phi_x^{i+1} \!= A\Phi_x^{i} \!+\! B\Phi_u^{i}, \Phi_x^{1} = I, \forall i, \label{eq:structure1}
\end{align}
\end{minipage}
\begin{minipage}[]{0.07\columnwidth}
\begin{equation*}
\left. \!\!\!\begin{array}{l} \\[1cm] \end{array} \right \rbrace\!,
\end{equation*}
\end{minipage}
\vspace*{-0.2cm}

imposing structure on the error system responses in an element-wise fashion. Constraint~\eqref{eq:structure2} enforces Toeplitz structure~\eqref{eq:Toeplitz-structure}, while constraint~\eqref{eq:structure1} is equivalent to the affine SLP constraint~\eqref{eq:SLP-affine-constraint}, given that the error system responses admit the Toeplitz structure. Additionally, the terminal ingredients $\mathcal{Z}_f$ and $P \in \mathbb{S}^n_{++}$ satisfy the following assumption.
\begin{assumption}\label{assump:terminal-ingredients}
Terminal ingredients $\mathcal{Z}_f, P$ are such that:
\begin{enumerate}
        \item[(i)] $\mathcal{Z}_f \subseteq \mathcal{X} \ominus \mathcal{F}_N^x$ is an RPI set with terminal control law $K_fx \in \mathcal{U} \ominus \mathcal{F}_N^u$, such that $\forall x \in \mathcal{Z}_f$ and $\forall w \in \mathcal{W}$, $(A+BK_f)x + \Gamma w \in \mathcal{Z}_f$, where we introduced the shorthand $\Gamma = A\Phi_{x^e}^{N} + B\Phi_{u^e}^{N}$.
        \item [(ii)] $\norm{x}^2_P$ is a Lyapunov function such that
        \begin{equation*}
            (A + BK_f)^\top P(A + BK_f) - P \preceq -Q - K_f^\top R K_f.
        \end{equation*}
\end{enumerate}
\end{assumption}
Since $\bm{\Phi}_\mathbf{x^e}, \, \bm{\Phi}_\mathbf{u^e}$ are optimization variables of~\eqref{SLTMPC:generic}, $\mathcal{F}_N^x$, $\mathcal{F}_N^u$, and $\Gamma$ are not available to compute $\mathcal{Z}_f$ offline. We discuss one possible approach to address this issue in Section~\ref{sec:secondary-process}. For more details on the SLP and SLTMPC, see~\citep{Anderson2019} and~\citep{Sieber2021}, respectively.
\begin{remark}
The set of structural constraints~$\mathcal{S}$ can be extended with additional constraints, like e.g., a finite impulse response constraint, i.e., $\Gamma = A\Phi_{x^e}^{N} + B\Phi_{u^e}^{N} = 0$, as in~\citep{Sieber2022}. In this case, $\mathcal{Z}_f$ reduces to a positively invariant set, since $\Gamma = 0$.
\end{remark}
SLTMPC has high computational complexity due to the concurrent optimization of the nominal trajectories $\mathbf{z},\,\mathbf{v}$ and the tube sequences $\mathcal{F}^x$, $\mathcal{F}^u$. To alleviate some of the complexity, one could fix~$\bm{\Phi}_\mathbf{x^e}, \,\bm{\Phi}_\mathbf{u^e}$ and therefore the tubes in~\eqref{SLTMPC:generic}, which results in an optimization problem of similar complexity to tube-based MPC at the expense of loosing some of the performance improvement, see~\citep{Sieber2022}. Instead of fixing a single tube sequence, the asynchronous computation mechanism proposed in this paper continuously computes new tube sequences in the secondary process. The computed tubes are passed to the primary process, which computes the nominal trajectory (see Figure~\ref{fig:asynch-up}). This separation decouples the computationally expensive tube computation from the computation of the nominal control input at the required sampling rate. Next, we first discuss the secondary process, before formulating the primary process and addressing the main theoretical challenge: ensuring recursive feasibility of the asynchronous computation mechanism.

\subsection{Secondary Process: Tube Optimization}\label{sec:secondary-process}
In general, any algorithm computing SL-DRS can be used as the secondary process, as long as the resulting error system responses obey structural constraint~\eqref{SLTMPC:structural-constraint} and the state and input constraints. Therefore, the secondary process can, e.g., solve the full SLTMPC problem~\eqref{SLTMPC:generic}. In this case, the terminal set cannot be computed offline, since $\mathcal{F}_{N}^x, \, \mathcal{F}_{N}^u$, and $\Gamma$ are only available online. Inspired by~\citep{Sieber2022}, where a scaled positively invariant set is used as the terminal set, we extend this idea to RPI sets. The goal is to find a scaling $\alpha \!\geq\! 0$, such that $\mathcal{Z}_f \!=\! \alpha \mathcal{X}_f$ is a valid terminal set for~\eqref{SLTMPC:generic}, where $\mathcal{X}_f$ is a fixed offline computed RPI set for system~\eqref{eq:dynamics} with terminal controller~$K_f$ and disturbance~$w \!\in\! \mathcal{W}$ subject to constraint~\eqref{eq:constraints}. Then, the scaled terminal set $\alpha\mathcal{X}_f$ needs to satisfy Assumption~\ref{assump:terminal-ingredients}.(i), i.e.,
\begin{subequations}\label{eq:terminal-set-conditions}
\begin{align}
\alpha (A+BK_f)\mathcal{X}_f &\subseteq \alpha \mathcal{X}_f \ominus \Gamma \mathcal{W}, \label{tsc:decrease}\\
\alpha \mathcal{X}_f &\subseteq \mathcal{X} \ominus \mathcal{F}_N^x, \label{tsc:state-inclusion}\\
\alpha K_f\mathcal{X}_f &\subseteq \mathcal{U} \ominus \mathcal{F}_N^u. \label{tsc:input-inclusion}
\end{align}
\end{subequations}
Using conditions~\eqref{eq:terminal-set-conditions}, the secondary process is defined as the SLTMPC with the scaled terminal set $\alpha\mathcal{X}_f$
\begin{subequations}\label{eq:process2-nominal}
        \begin{alignat}{2}
                \min_{\alpha, \mathbf{z}, \mathbf{v}, \bm{\Phi}_\mathbf{x^e}, \bm{\Phi}_\mathbf{u^e}} \quad & L(\mathbf{z}, \mathbf{v}, \bm{\Phi}_\mathbf{x^e}, \bm{\Phi}_\mathbf{u^e}) \\
                \textrm{s.t. } \qquad\;\, & \eqref{SLTMPC:dynamics} - \eqref{SLTMPC:init}, \\
                & \eqref{tsc:decrease} - \eqref{tsc:input-inclusion}, \label{process2-nominal:terminal-inclusion}
        \end{alignat}
\end{subequations}
where $L(\cdot)$ is a suitable cost function. A natural choice for $L(\cdot)$ is the nominal cost~\eqref{process1:cost}, however we can, e.g., also choose a cost minimizing the constraint tightening as in~\cite[Equation~(22)]{Sieber2022}, a performance cost like, e.g., $\mathcal{H}_\infty$ as shown in~\citep{Anderson2019}, or a regret optimal cost as proposed in~\citep{Didier2022}.
Note that it is always possible to find a scaled terminal set obeying~\eqref{process2-nominal:terminal-inclusion}, since the optimizer can set $\Gamma = 0$. Then, there exists a sufficiently small $\alpha$ such that~\eqref{tsc:state-inclusion} and~\eqref{tsc:input-inclusion} are satisfied, because $\alpha (A+BK_f)\mathcal{X}_f \subseteq \alpha\mathcal{X}_f$ holds for any $\alpha \geq 0$ by construction of $\mathcal{X}_f$. Furthermore, the terminal constraints~\eqref{process2-nominal:terminal-inclusion} can be imposed as linear constraints by applying the following lemma.
\begin{lem}\label{cor:suff-cond}
    For scalars $\alpha \geq 0$, $\beta \geq 0$, matrices $A \!\in\! \mathbb{R}^{p \times p}$, $\Gamma \in \mathbb{R}^{p \times p}$, polytopic sets $\mathcal{X} \!=\! \{a \mid H_x a \leq h_x, H_x \in \mathbb{R}^{n_x \times p} \}, \, \mathcal{Y} = \{a \mid H_y a \leq h_y, H_u \in \mathbb{R}^{n_y \times p}\}$, and compact convex set $\mathcal{Z} \subseteq \mathbb{R}^p$, the following conditions are sufficient for
    \begin{equation}\label{eq:cor}
	\alpha A \mathcal{X} \subseteq \beta \mathcal{Y} \ominus \Gamma \mathcal{Z}
	\end{equation}
	to hold:
	\begin{subequations}\label{eq:cor-suff-cond}
	\begin{align}
	&\exists \Lambda \in \mathbb{R}^{n_y \times n_x}, \, \Lambda \geq 0, \\
	&\Lambda H_x = \alpha H_y A, \\
	&\Lambda h_x \leq \beta h_y - h_{\mathcal{Z}}(\Gamma^\top H_y^\top),
	\end{align}
	\end{subequations}
	where $h_{\mathcal{Z}}(\Gamma^\top\! H_y^\top \!) \!=\! (h_\mathcal{Z}(\Gamma^\top\! H_y^{{1,:}^\top}), \dots, h_\mathcal{Z}(\Gamma^\top\! H_y^{{n_y,:}^\top}))$ denotes the stacked support functions.
\end{lem}
\vspace*{-0.2cm}
\begin{pf}
Using standard properties of polytopic sets, we define the scaled polytopes $\alpha \mathcal{X} = \{a \mid H_x a \leq \alpha h_x \}$ and $\beta \mathcal{Y} = \{a \mid H_y a \leq \beta h_y \}$. Then, using the fact that $\beta\mathcal{Y}$ is a polytope and Theorem~2.3 in~\citep{Kolmanovsky1998}, we get
\begin{equation*}
\beta\mathcal{Y} \ominus \Gamma \mathcal{Z} = \{ y \mid H_y y \leq \beta h_y - h_{\mathcal{Z}}(\Gamma^\top \! H_y^\top) \}.
\end{equation*}
Therefore,~\eqref{eq:cor} is a polytope containment problem, which can be restated using the following sufficient conditions from Theorem~1 in~\citep{Sadraddini2019}
\begin{align*}
&\exists \tilde{\Lambda} \in \mathbb{R}^{n_y \times n_x}, \, \tilde{\Lambda} \geq 0, \\
&\tilde{\Lambda}H_x = H_yA, \\
&\tilde{\Lambda}\alpha h_x \leq \beta h_y - h_{\mathcal{Z}}(\Gamma^\top H_y^\top).
\end{align*}
Finally, with the change of variables $\Lambda {=} \alpha \tilde{\Lambda}$, we get~\eqref{eq:cor-suff-cond}.\qed
\end{pf}
The secondary process~\eqref{eq:process2-nominal} computes the error system responses~$\bm{\Phi}_\mathbf{x^e}$, $\bm{\Phi}_\mathbf{u^e}$ that define the tubes $\mathcal{F}^x, \, \mathcal{F}^u$ and the terminal set~$\alpha\mathcal{X}_f$, asynchronously to the primary process. A finite number of these are then stored in memory $\mathtt{M} = \{ (\mathcal{F}_{j}^x,\, \mathcal{F}_{j}^u, \, \alpha_j) \mid j = 0,\dots,M\scalebox{0.9}[1.0]{\( - \)}1 \}$, where we denote the $j^\mathrm{th}$ memory entry as $\mathtt{M}_j$. A memory update is considered whenever the secondary process has computed a new solution~$\mathtt{M}_\mathrm{new} = (\mathcal{F}^x, \, \mathcal{F}^u, \, \alpha)$, rendering the memory time-varying. The update procedure is stated in Algorithm~\ref{alg:memory-update}. In order to retain feasibility when the memory is updated, the tubes and terminal sets in $\mathtt{M}$ are fused in the primary process using a convex combination with combination parameter $\bm{\lambda}(k) = (\lambda_0, \dots, \lambda_{M-1})$, similar to the robust MPC approach in~\citep{Kogel2020b}.

\subsection{Primary Process: Nominal Trajectory Optimization}
Using a convex combination of the tubes and terminal sets stored in memory at time $k$, i.e.~$\mathtt{M}(k)$, the primary process is defined as
\begin{subequations}\label{eq:process1}
        \begin{alignat}{2}
                \min_{\mathbf{z}, \mathbf{v}, \bm{\lambda}} \quad & \norm{z_N}^2_P + \sum_{i=0}^{N-1} \norm{z_i}^2_Q + \norm{v_i}^2_R + R(\bm{\lambda}) \label{process1:cost}\\
                \textrm{s.t. } \quad\! & \mathbf{z} = \mathbf{ZA}_{N+1}\mathbf{z} + \mathbf{ZB}_{N+1}\mathbf{v} + \mathbf{z}_\mathbf{0}, \\
                & z_i \!\in\! \bigoplus_{j=0}^{M-1} \lambda_j \left(\mathcal{X} \ominus  \mathcal{F}_{i,j}^x \right)\!, &&\hspace*{-1.5cm} i=0, \dots, N\!-\!1, \label{process1:state_constraints}\\
                & v_i \!\in\! \bigoplus_{j=0}^{M-1} \lambda_j \left(\mathcal{U} \ominus \mathcal{F}_{i,j}^u \right)\!, &&\hspace*{-1.5cm} i=0, \dots, N\!-\!1, \\
                & z_N \!\in\! \bigoplus_{j=0}^{M-1} \lambda_j \alpha_j \mathcal{X}_f, \label{process1:terminal-constraint} \\
                & \bm{\lambda} \geq 0, \  \bm{1}^\top \bm{\lambda} = 1, \\
                & z_0 = x(k),
        \end{alignat}
\end{subequations}
where $R(\bm{\lambda})$ is a regularization cost, which discourages the usage of older tubes by penalizing their convex combination weights, and $\mathcal{F}_{j}^x, \mathcal{F}_{j}^u, \alpha_j$ are the tube sequences and terminal set scaling stored in $\mathtt{M}_j$. Note that the terminal control law $K_f$ and the terminal cost $P$ are the same for all terminal sets, thus the sets $\alpha_j\mathcal{X}_{f}$ only differ in their size and shape, depending on the tightenings $\mathcal{F}_{N,j}^x,\, \mathcal{F}_{N,j}^u$ and on $\Gamma_j$. The resulting MPC control law is given by~$\kappa(x(k)) = v_0^*$, where $v_0^*$ is the first element of an optimizer $\mathbf{v}^*$ of~\eqref{eq:process1}.
\begin{algorithm}[h]
\caption{Update memory $\mathtt{M}$}\label{alg:memory-update}
\hspace*{\algorithmicindent}\! \textbf{Input:} $\mathtt{M}_\mathrm{new}$, $\mathtt{M}(k\scalebox{0.9}[1.0]{\( - \)}1)$, $\bm{\lambda}(k\scalebox{0.9}[1.0]{\( - \)}1)$ \\
\hspace*{\algorithmicindent}\! \textbf{Output:} $\mathtt{M}(k)$
\begin{algorithmic}[1]
\Procedure{updateMemory}{}
\State $\mathtt{M}(k) \gets \mathtt{M}(k\scalebox{0.9}[1.0]{\( - \)}1)$
\If{$\mathtt{M}(k)$ not full}
	\State $\mathtt{M}_j(k) \gets \mathtt{M}_\mathrm{new}$, where $\mathtt{M}_{j}(k)$ is empty
\ElsIf{$\exists j \ \mathrm{s.t.} \ \lambda_{j}(k\scalebox{0.9}[1.0]{\( - \)}1) = 0$}
	\State \(\triangleright\) $\mathtt{M}_{j}(k)$ does not contribute to convex comb.
    \State $\mathtt{M}_j(k) \gets \mathtt{M}_\mathrm{new}$, where $j$ s.t. $\lambda_{j}(k\scalebox{0.9}[1.0]{\( - \)}1) = 0$
\Else
	\State discard $\mathtt{M}_\mathrm{new}$
\EndIf
\EndProcedure
\end{algorithmic}
\end{algorithm}

\subsection{Theoretical Guarantees}
In the following, we provide theoretical guarantees for the proposed SLTMPC with asynchronous computation, i.e., we prove recursive feasibility of~\eqref{eq:process1}, including the case in which the memory is updated. For this, we make the following assumption.
\begin{assumption}\label{assump:init-feas}
At the start of the control task,~$\mathtt{M}$ contains at least one entry for which the primary process is feasible, i.e.,~\eqref{eq:process1} is feasible for $\mathtt{M}_0(0) = (\mathcal{F}_{0}^x, \, \mathcal{F}_{0}^u, \, \alpha_0)$ with $\lambda_0(0) = 1$.
\end{assumption}
Assumption~\ref{assump:init-feas} is not restrictive since we can always compute a sequence of tubes offline, e.g., by computing the DRS. We show recursive feasibility in a similar fashion to~\citep{Kogel2020a}, but consider the more general case of SL-DRS tubes instead of DRS tubes.
\begin{prop}
    Let Assumption~\ref{assump:init-feas} hold. Then, primary process~\eqref{eq:process1} is recursively feasible for any constant memory~$\mathtt{M}(k) = \mathtt{M}$. Furthermore, using the time-varying memory~$\mathtt{M}(k)$ updated by Algorithm~\ref{alg:memory-update} maintains recursive feasibility.
\end{prop}
\begin{pf}
We first prove that~\eqref{eq:process1} is recursively feasible for any convex combination of constant memory entries $\mathtt{M}_j = (\mathcal{F}_{j}^x, \, \mathcal{F}_{j}^u, \, \alpha_j)$, before showing that the memory can be updated using Algorithm~\ref{alg:memory-update}. Let $\mathbf{z}^*(x(k)) = (z_0^*, \dots, z_N^*)$, $\mathbf{v}^*(x(k)) = (v_0^*, \dots, v_{N-1}^*)$, and $\bm{\lambda}^*(k) = (\lambda_0^*, \dots, \lambda_{M-1}^*)$ denote the optimal solution of~\eqref{eq:process1} for initial state $x(k)$ and let $x^+ = x(k\!+\!1)$ be the system state at the next time step. Then, we construct the candidate sequences for initial state $x^+$ using a standard shifting approach as $\hat{\bm{\lambda}}(k\scalebox{0.9}[1.0]{\( + \)}1) = \bm{\lambda}^*(k)$, $\hat{\mathbf{z}}(x^+) = (\hat{z}_0, \dots, \hat{z}_N)$, and $\hat{\mathbf{v}}(x^+) = (\hat{v}_0, \dots, \hat{v}_{N-1})$, where
\begin{subequations}\label{pf:candidates}
\begin{alignat}{2}
\hat{z}_i &= z_{i+1}^* + \sum_{j=0}^{M-1} \lambda_j^* \Phi_{x^e,j}^{i+1} w(k), &&\hspace*{-1.5cm}\quad i=0,\dots,N\!-\!1,\\
\hat{z}_N &= \sum_{j=0}^{M-1} \lambda_j^*(Az_N^* +B\kappa_f(z_N^*)+ \Gamma_j w(k)), \\
\hat{v}_i &= v_{i+1}^* + \sum_{j=0}^{M-1} \lambda_j^* \Phi_{u^e,j}^{i+1} w(k), &&\hspace*{-1.5cm}\quad i=0,\dots,N\!-\!1,
\end{alignat}
\end{subequations}
with $\Gamma_j = A\Phi_{x,j}^N + B\Phi_{u,j}^N$.
Next, we show that these candidates fulfill the nominal dynamics. The state candidate sequence is initialized with
\begin{equation*}
\hat{z}_0 = x(k+1) = Ax(k) + Bv_0^* + w(k) = z_1^* + w(k)
\end{equation*}
and for $i=0, \dots, N-1$, it holds that
\begin{align*}
\hat{z}_{i+1} &= A\hat{z}_i + B\hat{v}_i
= z_{i+2}^* +\! \sum_{j=0}^{M-1} \lambda_j^* \left[ A\Phi_{x^e,j}^{i+1} \!+\! B\Phi_{u^e,j}^{i+1}\right] w(k) \\
&\overset{\mathclap{\strut\text{\eqref{eq:structure1}}}}= z_{i+2}^* +\! \sum_{j=0}^{M-1} \lambda_j^* \Phi_{x^e,j}^{i+2} w(k).
\end{align*}
For the last state of the state candidate, we have
\begin{align*}
\hat{z}_{N} &= A\hat{z}_{N-1} + B\hat{v}_{N-1} \\
&= Az_{N}^* + B\kappa_f(z_{N}^*) + \sum_{j=0}^{M-1} \lambda_j^* \left[ A\Phi_{x^e,j}^{N} + B\Phi_{u^e,j}^{N}\right] w(k) \\
&= \sum_{j=0}^{M-1} \lambda_j^* \left( Az_{N}^* + B\kappa_f(z_{N}^*) +\Gamma_j w(k)\right),
\end{align*}
where we used $\sum_{j=0}^{M-1} \lambda_j^* = 1$ and the definition of $\Gamma_j$. Using $\sum_{j=0}^{M-1} \lambda_j^* = 1$ again, we rewrite the state and input candidates as
\begin{align*}
\hat{z}_i \!=\!\! \sum_{j=0}^{M-1} \!\!\lambda_j^* \!\left( z_{i+1}^* \!+\!  \Phi_{x^e,j}^{i+1} w(k) \!\right)\!, \,
\hat{v}_i \!=\!\! \sum_{j=0}^{M-1} \!\!\lambda_j^* \!\left( v_{i+1}^* \!+\! \Phi_{u^e,j}^{i+1} w(k) \!\right)\!.
\end{align*}
Then, we need to show that the candidates fulfill
\begin{align}
z_{i+1}^* +  \Phi_{x^e,j}^{i+1} w(k) \in \mathcal{X} \ominus \mathcal{F}_{i,j}^x, \label{proof:state_containment}\\
v_{i+1}^* + \Phi_{u^e,j}^{i+1} w(k) \in \mathcal{U} \ominus \mathcal{F}_{i,j}^u. \label{proof:input_containment}
\end{align}
Since $z_{i+1}^* \in \mathcal{X} \ominus \mathcal{F}_{i+1,j}^x$ and $\Phi_{x^e,j}^{i+1} w(k) \in \Phi_{x^e,j}^{i+1}\mathcal{W}$, we can show that~\eqref{proof:state_containment} holds with
\begin{align}
&\left(\mathcal{X} \!\ominus\! \mathcal{F}_{i+1,j}^x \right) \oplus \Phi_{x^e,j}^{i+1}\mathcal{W} \subseteq \mathcal{X} \ominus \mathcal{F}_{i,j}^x, \label{proof:suff-state-cont}
\end{align}
where we use~\cite[Theorem~2.1~(ii,v)]{Kolmanovsky1998} and~\eqref{SLDRS:inclusion}. Similarly,~\eqref{proof:input_containment} holds since
\begin{align}
&\left(\mathcal{U} \ominus \mathcal{F}_{i+1,j}^u \right) \oplus \Phi_{u^e,j}^{i+1}\mathcal{W} \subseteq \mathcal{U} \ominus \mathcal{F}_{i,j}^u. \label{proof:suff-input-cont}
\end{align}
We have shown~\eqref{proof:state_containment} and~\eqref{proof:input_containment}, therefore
$
\hat{z}_i \in \bigoplus_{j=0}^{M-1} \lambda_j^* ( \mathcal{X} \ominus \mathcal{F}_{i,j}^x ), \,
\hat{v}_i \in \bigoplus_{j=0}^{M-1} \lambda_j^* ( \mathcal{U} \ominus \mathcal{F}_{i,j}^u )$,
as claimed. For terminal constraint~\eqref{process1:terminal-constraint}, we get
$
\hat{z}_N \in \bigoplus_{j=0}^{M-1} \lambda_j^* \alpha_j\mathcal{X}_{f}
$
if $z_N^* \in \alpha_j\mathcal{X}_{f}$, which is true by definition of $\mathtt{M}_j$. It remains to show that we retain recursive feasibility when using $\mathtt{M}(k)$. Since Algorithm~\ref{alg:memory-update} only updates the entry $\mathtt{M}_{j}(k)$ if the corresponding $\lambda_{j}(k) = 0$, recursive feasibility is retained by keeping $\hat{\lambda}_{j}(k+1) = 0$ for $\mathtt{M}_{j}(k+1)$.\vspace*{-0.57cm}\flushright\qed
\end{pf}

\section{Numerical Results}\label{sec:numerical_section}
\begin{figure}[h!]
\centering
\includegraphics[width=0.89\columnwidth]{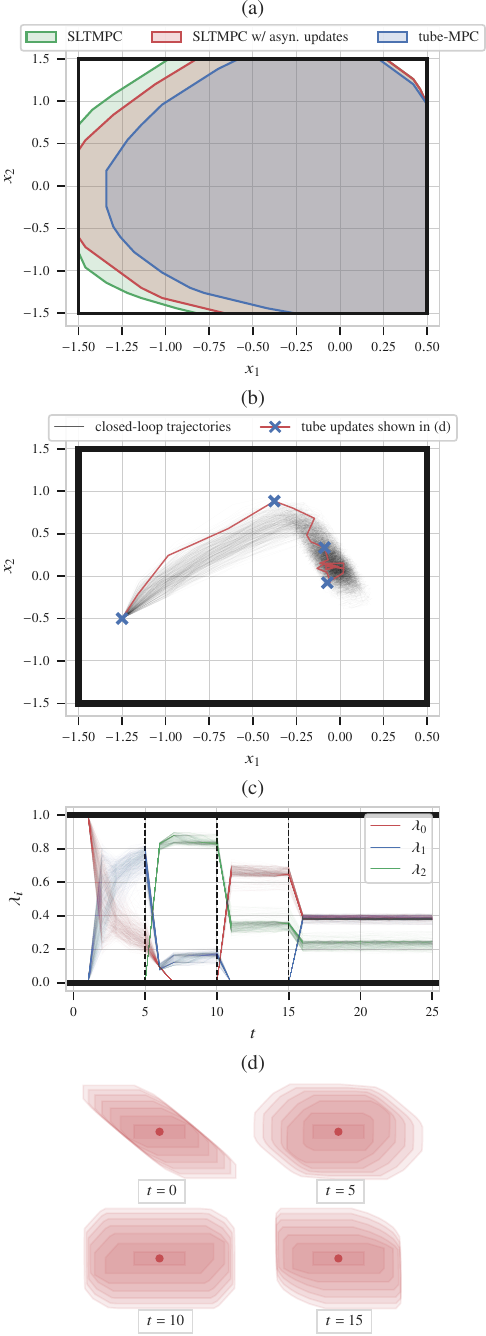}
\caption{(a): Comparison of SLTMPC, tube-MPC, and SLTMPC with asynchronous computation with respect to their RoA, (b-c): closed-loop behavior of~\eqref{eq:process1} with memory updated every fifth time-step (dashed lines in (c)) for initial state $x(0) = (-1.25, -0.5)$, and (d): shape of tubes for highlighted trajectory in (b).}\label{fig:NE}
\end{figure}

In the following, we compare SLTMPC with asynchronous computation to SLTMPC~\eqref{SLTMPC:generic} and tube-MPC~\citep{Chisci2001}. Additionally, we show how the tubes are updated over time using the asynchronous computation mechanism. All examples are implemented in Python using CVXPY~\citep{cvxpy}, are solved using MOSEK~\citep{mosek}, and are run on a machine equipped with an Intel~i7-8665U~(\unit[1.9]{GHz}) CPU. In all experiments, we consider uncertain system~\eqref{eq:dynamics} with discrete-time dynamic matrices
\begin{equation*}
A = \begin{bmatrix} 1.05 & 0.25 \\ 0 & 1 \end{bmatrix}, \quad B = \begin{bmatrix} 0.5 \\ 0.5 \end{bmatrix},
\end{equation*}
subject to the polytopic constraints
\begin{equation*}
\begin{bmatrix} \scalebox{0.7}[1.0]{\( - \)}1.5 \\ \scalebox{0.7}[1.0]{\( - \)}1.5 \end{bmatrix} \!\!\leq\!\! \begin{bmatrix} x_1 \\ x_2 \end{bmatrix} \!\!\leq\!\! \begin{bmatrix} 0.5 \\ 1.5 \end{bmatrix}\!, \scalebox{0.7}[1.0]{\( - \)}0.75 \!\leq\! u \!\leq\! 0.75, \begin{bmatrix} \scalebox{0.7}[1.0]{\( - \)}0.1 \\ \scalebox{0.7}[1.0]{\( - \)}0.1 \end{bmatrix} \!\!\leq\!\! \begin{bmatrix} w_1 \\ w_2 \end{bmatrix} \!\!\leq\!\! \begin{bmatrix} 0.1 \\ 0.1 \end{bmatrix}\!,
\end{equation*}
cost parameters $Q=10\cdot I$, $R=1$, and horizon $N=8$. For all three control methods, we use the LQR gain - computed using the same $Q$ and $R$ matrices - as terminal controller $K_f$ and compute terminal set $\mathcal{X}_f$ as the minimal RPI set using~$K_f$. \\
Figure~\ref{fig:NE}(a) shows the sets of feasible initial states (RoA) for SLTMPC, tube-MPC, and SLTMPC with asynchronous computation, where we use a memory of size $M=2$ and initialize the memory with the DRS~($\mathtt{M}_0$) and a tube sequence computed by~\eqref{SLTMPC:generic} for $x_0 = (-1, 0)$~($\mathtt{M}_1$). Combining the DRS with an additional sequence of tubes substantially increases the RoA compared to tube-MPC, which shows the reduction in conservativeness due to updating the tubes online. Figures~\ref{fig:NE}(b) and~\ref{fig:NE}(c) show $500$ closed-loop trajectories starting in $x(0) = (-1.25, -0.5)$ for SLTMPC with asynchronous computation and the corresponding convex combination variables $\lambda_j$, respectively, over $25$ time steps. The memory ($M=3$) is initialized with the DRS~($\mathtt{M}_0$) and tubes computed by~\eqref{eq:process2-nominal} with an $\mathcal{H}_\infty$-cost~($\mathtt{M}_1$). Then, every fifth time-step the memory is updated with tubes computed by~\eqref{eq:process2-nominal} using nominal cost~\eqref{process1:cost}. To further highlight the updated tubes, Figure~\ref{fig:NE}(d) shows the tubes at different time steps for the highlighted closed-loop trajectory in Figure~\ref{fig:NE}(b). Initially, only the DRS tubes are used ($\lambda_0$ in Fig.~\ref{fig:NE}(c)), before the $\mathcal{H}_\infty$-tubes become dominant ($\lambda_1$ in Fig.~\ref{fig:NE}(c)), which is also apparent in Figure~\ref{fig:NE}(d) when comparing time steps $t=0$ (only DRS) and $t=5$ (combination of DRS and $\mathcal{H}_\infty$-tubes). At time steps $t=5$ and $t=10$ new tube sequences are added, which immediately contribute heavily ($\lambda_2$ and $\lambda_0$, respectively, as well as tubes at $t=10$ and $t=15$, respectively). Since $\lambda_2 = 0$ at $t=5$, $\lambda_0 = 0$ at $t=10$, and $\lambda_1 = 0$ at $t=15$, the old tubes are successfully replaced by new ones (indicated with dashed lines). However at $t=20$, $\lambda_j \neq 0$ for any $j$, hence the tubes are not updated, which happens because the trajectories are close to the origin and updating the tubes is not beneficial anymore. As stated in Table~\ref{table:comp-times}, the minimum solve time for~\eqref{eq:process1} is \unit[1.74]{ms}, which is comparable to tube-MPC with \unit[1.52]{ms}, while SLTMPC~\eqref{SLTMPC:generic} and~\eqref{eq:process2-nominal} are significantly slower with \unit[19.32]{ms} and \unit[19.19]{ms}, respectively; and the average closed-loop cost for SLTMPC with asynchronous computation~($67.36$) lies between the closed-loop costs of SLTMPC~($59.07$) and tube-MPC~($85.92$), respectively.
\begin{table}[h]
\centering
\caption{Comparison of closed-loop costs and computation times for $500$ noisy trajectories starting in $x(0) = (-1.25, -0.5)$.}\label{table:comp-times}
\begin{tabular}{@{}lcccc@{}}
\toprule
& \multicolumn{2}{c}{Cost [-]} & \multicolumn{2}{c}{Comp. Time [ms]} \\ \cmidrule(lr){2-3} \cmidrule(lr){4-5}
& Mean & Std. Dev. & Min. & Median \\ \midrule
Primary Process~\eqref{eq:process1} & 67.36 & 11.23 & 1.74 & 3.04\\
Secondary Process~\eqref{eq:process2-nominal} & N/A & N/A & 19.19 & 38.92\\
SLTMPC~\eqref{SLTMPC:generic} & 59.07 & 9.87 & 19.32 & 33.94\\
tube-MPC & 85.92 & 14.35 & 1.52 & 2.07\\
\bottomrule 
\end{tabular}
\end{table}

\clearpage
\section{Conclusions}\label{sec:conclusions}
This paper introduced an asynchronous computation mechanism for system level tube-MPC (SLTMPC) applied to constrained linear time-invariant systems with additive disturbances. By splitting the nominal trajectory and the tube computations into separate processes - a primary and a secondary process - the computational complexity of SLTMPC is significantly reduced. We proved recursive feasibility of the proposed method allowing that a set of tube sequences can be stored and updated in the memory of the primary process. Finally, we showed the benefits and behavior of the proposed method for a numerical example.

\section*{Data Availability Statement}
All data and figures generated for this paper are available on the ETH Research Collection under \url{doi.org/10.3929/ethz-b-000606960}.

\balance


\bibliography{bibliography}


\end{document}

%% file: tikz/comparison.tex

\begin{tikzpicture}[>=latex,scale = 0.85]

\draw[thick]  (6.7,-2) rectangle (0.7,0);
\node[align = center] at (3.7,-0.5) {\textbf{Tube-based MPC}};
\node[align = left] at (3.7,-1.3) {
{\small-~Nominal Trajectory (online)}\\
{\small-~Tubes (offline)}
};
\draw[-latex,thick] (-0.1,-1) -- (0.7,-1);
\node at (0.3,-0.75) {$x$};
\draw[-latex,thick] (6.7,-1) -- (7.5,-1);
\node at (7.1,-0.75) {$u$};

\draw[thick]  (6.7,-4.2) rectangle (0.7,-2.2);
\node[align = center] at (3.7,-2.7) {\textbf{SLTMPC}};
\node[align = left] at (3.7,-3.5) {
{\small-~Nominal Trajectory (online)}\\
{\small-~Tubes (online)}
};
\draw[-latex,thick] (-0.1,-3.2) -- (0.7,-3.2);
\node at (0.3,-2.95) {$x$};
\draw[-latex,thick] (6.7,-3.2) -- (7.5,-3.2);
\node at (7.1,-2.95) {$u$};

\draw[thick]  (6.7,-9.3) rectangle (0.7,-4.4);
\node[align = center] at (3.7,-4.9) {\textbf{Asynchronous Computation}};
\draw[-] (0.7,-6.3) -- (1.5,-6.3);
\draw[-] (1.1,-6.3) -- (1.1,-8.5) -- (1.5,-8.5);
\draw[dashed]  (6.,-7.3) rectangle (1.5,-5.3);
\node[align = center] at (3.75,-5.65) {\textit{Primary Process}};
\node[align = left] at (3.75,-6.55) {
{\small-~Nominal Trajectory}\\ \quad\!\!{\small (online)}\\
{\small-~Use Tubes $\mathcal{F}$}
};
\draw[latex-] (3.7,-7.3) -- (3.7,-7.9);
\node at (3.4,-7.65) {\small$\mathcal{F}$};
\draw[dashed]  (6.,-9.1) rectangle (1.5,-7.9);
\node[align = center] at (3.75,-8.25) {\textit{Secondary Process}};
\node[align = left] at (3.75,-8.9) {
{\small-~Tubes $\mathcal{F}$ (online)}\\
};
\draw[-] (6.0,-6.3) -- (6.7,-6.3);
\draw[-latex,thick] (-0.1,-6.3) -- (0.7,-6.3);
\node at (0.3,-6.1) {$x$};
\draw[-latex,thick] (6.7,-6.3) -- (7.5,-6.3);
\node at (7.1,-6.1) {$u$};

\end{tikzpicture}

%% file: tikz/asynchronous-updates.tex

\begin{tikzpicture}[>=latex,scale = 0.95]

\draw[-latex,thick] (0,-0.6) -- (4.6,-0.6);
\node at (0.1,-0.4) {$x$};
\draw[-latex,thick] (0.25,-0.6) -- (0.25,-1.55) -- (0.6,-1.55);

\draw[thick]  (7.,-2.2) rectangle (4.6,0);
\node[align = center] at (5.8,-0.6) {Primary\\Process~\eqref{eq:process1}};

\draw[dashed]  (6.8,-2.) rectangle (4.8,-1.1);
\node[align = center] at (5.8,-1.55) {{\small Memory $\mathtt{M}$}\\{\small (\textit{Size:} $M$)}};

\draw[-latex, thick] (3.1,-1.55) -- (4.6,-1.55);
\node at (3.85,-1.3) {$\mathcal{F}^x, \mathcal{F}^u$};
\node at (3.85,-1.8) {$\alpha\mathcal{X}_f$};

\draw[thick]  (3.1,-2.1) rectangle (0.6,-1);
\node[align = center] at (1.85,-1.55) {Secondary\\Process~\eqref{eq:process2-nominal}};

\draw[-latex,thick] (7,-0.6) -- (7.8,-0.6);
\node at (7.4,-0.4) {$u$};






\end{tikzpicture}